\documentclass[aps,prb,twocolumn,superscriptaddress]{revtex4}
\usepackage{graphicx}

\begin{document}

\title{
Topological order in the insulating Josephson junction array.
} 

\author{B. Dou\c{c}ot}
\affiliation{Laboratoire de Physique Th\'{e}orique et Hautes 
\'Energies, CNRS UMR 7589,
Universit\'{e}s Paris 6 et 7, 4, place Jussieu, 
75252 Paris Cedex 05 France}

\author{M.V. Feigel'man}
\affiliation{Landau Institute for Theoretical Physics, Kosygina 2, Moscow, 117940 Russia}

\author{L.B. Ioffe} 

\altaffiliation{Landau Institute for Theoretical Physics, Kosygina 2, Moscow, 117940 Russia}

\affiliation{Center for Materials Theory,  
Department of Physics and Astronomy, Rutgers University 
136 Frelinghuysen Rd, Piscataway NJ 08854 USA}

\begin{abstract}

We propose a Josephson junction array which can be tuned into an unconventional 
insulating state by varying external magnetic field. This insulating state 
retains a gap to half vortices; as a consequence, such array with non-trivial
global geometry exhibits a ground state degeneracy. This degeneracy is
protected from the effects of external noise. We compute the gaps
separating higher energy states from the degenerate ground state and 
we discuss experiments probing the unusual properties of this insulator. 

\end{abstract}

\maketitle


A device implementing an ideal quantum computer would be a very
interesting object from a physics point of view: it is a system with
exponentially many degenerate states ($\mathcal{N\sim }2^{K}$, $K\sim 10^{4}$%
) and extremely low decoherence rate \cite{Preskill1998}. The latter implies
that the coupling of these states to the noise field produced by environment
should be very small, more precisely all states are coupled to the external
noise in exactly the same way:  $\langle n\left| \widehat{O}\right| m\rangle
=O_{0}\delta_{mn} + o(e^{-L})$ where $\widehat{O}$ is an operator of the physical 
noise and $L$ is a parameter such as the system size that can be made as large as
desired. This condition is naturally satisfied \cite{Kitaev1997} if
degenerate states are distinguished only by a non-local topological order
parameter. \cite{Wen1990,Wen1991}. In the previous work \cite
{Ioffe2002a} we have proposed a superconducting Josephson junction array
that, in addition to superconducting long range order, acquires a
topological order parameter for the arrays with non-trivial geometry.  The
goal of this paper is to identify the Josephson junction array with the
ground state characterized \emph{only} by the topological order parameter
and no other long range order of any other kind (in local variables)
and which has a gap to all other (non-topological) excitations. 
Because such array has neither
superconducting phase stiffness nor gapless excitations it can not carry an
electric current; thus in the following we shall call it topological
insulator. While the arrays with similar global properties were proposed and
studied in a recent work  \cite{Ioffe2002b}, the arrays discussed here have
a number of important practical advantages which 
makes them more feasible to built and measure in laboratory.

In physical terms, the topological superconductor appearing in the array 
\cite{Ioffe2002a} is a superfluid of $4e$ composite objects.  
The topological degeneracy of the ground state arises 
because $2e$ excitations have a gap. Indeed, in such system with the geometry 
of an annulus, one extra Cooper pair injected at the inner boundary can never
escape it; on the other hand, it is clear that two states differing by the 
parity of the number of Cooper pairs at the boundary are practically 
indistinguishable by a local measurement.

Generally, increasing the charging energy in a Josephson junction array makes 
it an insulator. This transition is due to an increase of phase fluctuations 
in the original array and the resulting appearance of free
vortices that form a superfluid of their own. The new situation arises
in topological superconductor because it allows half-vortices. 
Two scenarios are now possible. 
The "conventional" scenario would involve condensation of half-vortices
since they are conjugate to $4e$ charges. In this case we get an insulator
with elementary excitations carrying charge $4e$. An alternative is 
condensation of full vortices (pairs of half vortices) with a finite
gap to half vortices. In this case the elementary excitations are
charge $2e$ objects. Similar fractionalization was discussed in the 
context of high $T_c$ superconductors in \cite{Senthil2000,Senthil2001} 
and in the context of spin or quantum dimer systems in 
\cite{Sachdev1991,Moessner2002,Balents2002,Senthil2002,Misguich2002}.
Such insulator acquires interesting topological properties on a 
lattice with holes because each hole leads to a new binary degree
of freedom which describes presence or absence of half vortex.
The energies of these states are equal up to corrections which
vanish exponentially with the size of the holes. These states
cannot be distinguished by local measurements and have all 
properties expected for a topological insulator. They can be
measured, however, if the system is adiabatically
brought into the superconductive state by changing some controlling
parameter. In this paper we propose a modification of the array
\cite{Ioffe2002a} that provides such control parameter and, at the
same time, allows us to solve the model and compute the properties
of the topological insulator. 
 


The main physical idea of the array is to give a system where
kinetic energy of half vortex, $t_{hv}$ is \textit{parametrically} smaller than the kinetic
energy of the full vortex,  $t_{fv}$ and where their potential energies, $W$,
satisfy $t_{hv}<W<t_{fv}$. 
The array is shown in Fig.~1, it contains rhombi with junctions
characterized by Josephson and charging energies $E_J>E_C$ 
and weak junctions with $\epsilon _{J} \ll \epsilon _{C} \ll E_C$. 
Each rhombus encloses half of
a flux quantum leading to an exact degeneracy between the two states of 
opposite chirality of the circulating current \cite{Ioffe2002a,Doucot2002}. 
This degeneracy is a consequence of the symmetry operation which combines 
the reflection about the long diagonal of the rhombus and a gauge transformation 
needed to compensate the change of the flux $\Phi_0/2 \rightarrow -\Phi_0/2$. 
This gauge transformation changes the phase difference along the 
diagonal by $\pi$. This ${\cal Z}_2$ symmetry implies the conservation
of the parity of the number of pairs at each site of the hexagonal lattice
and is the origin of the Cooper pair binding. 
We assume that each elementary hexagon contains exactly $k$ 
weak junctions: in case each link contains one weak junction $k=6$, but 
generally it can take any value $k \geq 1$. As will be shown below, the 
important condition is the number of weak junctions that one needs to cross 
in the elementary loop. Qualitatively, a value $k \geq 1$ ensures that it costs 
a little to put vortex in any hexagon.

For the general arguments that follow below the actual construction of the
weak links is not important, however, practically it is difficult to vary
the ratio of the capacitance to the Josephson energy so weaker Josephson
contact usually implies larger Coulomb energy. This can be avoided if weak
contact is made from Josephson junction loop frustrated by magnetic
field. The charging energy of this system is half the charging energy of
the individual junction while the effective Josephson junction strength is
$\epsilon_J=2\pi \frac{\delta \Phi}{\Phi_0} E_0$ where $E_0$ is the Josephson energy of
each contact and $\delta \Phi=\Phi-\frac{\Phi_0}{2}$ is the difference of the flux
from half flux quanta. This construction also allows
to control the system by varying the magnetic field. 

Assume that $\epsilon _{J}$ sets the lowest energy scale in this
problem (the exact condition will be discussed below). The state of the
array is controlled by discrete variables $u_{ab}=0,1$ which describe the
chiral state of each rhombus  and by continuous phases $\phi _{ab}$ that 
specify the state of each weak link (here and below $a,b$ denote the sites
of hexagonal lattice). If Josephson coupling $\epsilon _{J}\equiv 0$
different islands are completely decoupled and potential energy does not
depend on discrete variables $u_{ab}$. For small $\epsilon _{J}$ we can
evaluate its effect in the perturbation theory: 
\begin{equation}
V(u)=-W\cos (\pi \sum_{hex}u_{ab}),\;
W= \frac{k^k}{k!} 
\left(\frac{\epsilon _{J}}{8\epsilon _{C}}\right) ^{k-1}  \epsilon _{J} \label{V(u)}
\end{equation}
This potential energy lowers the
energy of classical configurations of $u_{ab}$ that
satisfy the constraint $\sum_{hex}u \equiv 0\;[2]$ but does not prohibit
the ones with $\sum_{hex}u \equiv 1 \;[2]$.

\begin{figure}[ht]
\includegraphics[width=3.2in]{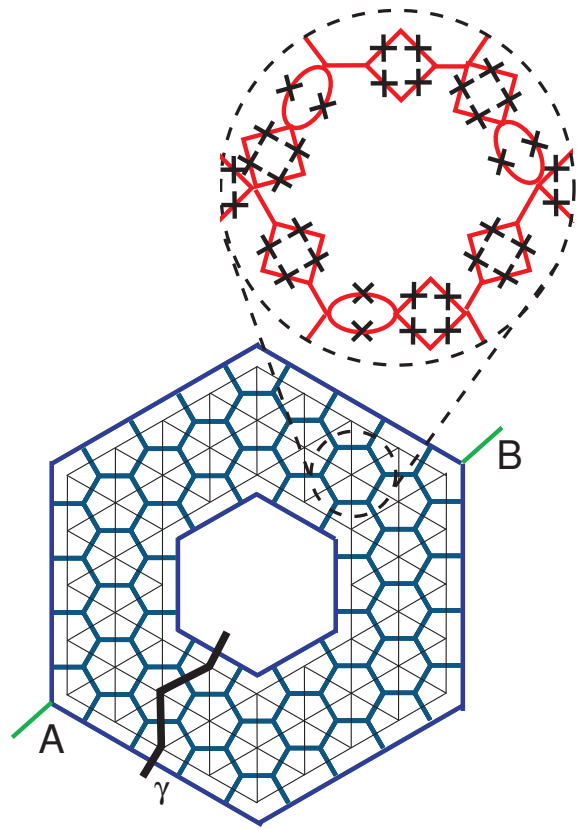}
\caption{
Schematics of the array. The main figure: global structure of the array. 
Discrete variables controlling the low energy properties are defined on 
the links of the hexagonal lattice. Generally, the lattice might have 
$K$ big holes, here we show $K=1$ example. Zoom in: Each inner
bond of the lattice contains a rhombus made out of four Josephson junctions;
some bonds also contain an effective weaker link made of two Josephson junctions
so that each hexagon of the lattice contains $k=3$ such links. 
The flux through each rhombus is half-flux quanta, $\Phi_0/2$, 
the flux through a loop constituting a weak link is close to half flux quanta 
$\Phi=\Phi_0/2+\delta\Phi$. The boundary of the lattice contains rhombi and
weak links so that each boundary plaquette has the same number, $k$,
of weak links as the bulk hexagon. 
}
\end{figure}

Consider now the dynamics of discrete variables. Generally, two types of
tunneling processes are possible. In the first type the phase changes by $%
\pi $ across each of the three rhombi that have a common site. This is the
same process that gives the leading contribution to the dynamics of the
superconducting array \cite{Ioffe2002a}, its amplitude is given 
(in the quasiclassical approximation) by 
\begin{equation}
r\approx E_{J}^{3/4}E_{C}^{1/4}\exp (-3S_{0}),\;\;S_{0}=1.61\sqrt{E_{J}/E_{C}%
}  \label{r}
\end{equation}
In the second type of process the phase changes across one rhombus and
across one weak junction. Because the potential energy of the weak junction
is assumed to be very small the main effect of the weak junction is to
change the kinetic energy. The total kinetic energy for this process is
the sum of the terms due to the phase across the rhombi and across the weak
link. Assuming that these phase variations are equal and opposite in sign, the 
former is about $E_C^{-1} \dot{\phi}^2$ while the latter 
$\epsilon_C^{-1} \dot{\phi}^2$, so the effective charging energy of this process 
is 
$
\widetilde{E_{C}}=\left( E_{C}^{-1}+\epsilon _{C}^{-1} \right)^{-1}
$.
For $\epsilon _{C}\ll E_{C}$ this charging energy is small and such process
is suppressed. Thus, in these conditions the dominating process is the
simultaneous flip of three rhombi as in the superconducting case. In the
following we restrict ourselves to this case. Further, we shall assume that $%
r \gg W$ so that in the leading order one can neglect the
potential energy compared to the kinetic energy corresponding to the flip of
three rhombi. 
As $W$ is increased by turning on $\epsilon_J$ the continuous phase $\phi_{ab}$
orders and the transition into the superconducting state happens at 
$\epsilon_J \sim \epsilon_C$. At larger
$\epsilon_J$, $W$ becomes $\epsilon_J$ and with a further increase of 
$\epsilon_J$, for $\epsilon_J \gg r$ vortices completely disappear from 
the low energy spectrum and the array becomes equivalent to the one studied 
in \cite{Ioffe2002a}.   

The low energy states are the ones that minimize the kinetic energy
corresponding to simultaneous flip processes: 
\begin{equation}
H_{T}=-r\sum_{a}\prod_{b(a)}\tau_{ab}^x  \label{H_T}
\end{equation}
Here $b(a)$ denote the nearest neighbors of site $a$, $\tau_{ab}^x$
is the operator that flips discrete variables $u_{ab}$
and $r$ is given by (\ref{r}). The states minimizing this energy satisfy
the gauge invariance condition 
\begin{equation}
\prod_{b(a)}\tau_{ab}^x|\Psi \rangle =|\Psi \rangle  
\label{Gauge_inv}
\end{equation}
The Hilbert space of states that satisfy the condition (\ref{Gauge_inv}) is
still huge. If all weaker terms in the Hamiltonian are neglected all states
that satisfy (\ref{Gauge_inv}) are degenerate. These states can be visualized
in terms of half vortices positioned on the sites of the dual lattice, $i,j$.  
Indeed, a convenient way to describe different states that satisfy 
(\ref{Gauge_inv}) is to note that operator $\prod_{b(a)}\tau_{ab}^x$ 
does not change the value of $\sum_{j(i)}u \; [2]$. Thus, one can fix the 
values of $\sum_{j(i)}u_{ij}=v_{i}$ on all plaquettes, $i$ and impose 
the constraint (\ref{Gauge_inv}). In physical terms the binary values 
$v_{i}=0,1$ describe the positions of half-vortices on dual lattice.
This degeneracy between different states is lifted
when the subdominant terms are taken into account.   
The main contribution to the potential energy
of these half-vortices comes from (\ref{V(u)}), it is simply proportional to
their number. The dynamics of these vortices is due to the processes in
which only one rhombus changes its state and the corresponding flip of the
phase accross the weak junction. The amplitude of this process is

\[
\widetilde{r}\approx E_{J}^{3/4} \tilde{E}_{C}^{1/4}
\exp (-\widetilde{S_{0}}),\;\;
\widetilde{S_{0}}=1.61\sqrt{\frac{E_{J}}{\widetilde{E_{C}}}} 
\]

The effective Hamiltonian of half-vortices is
\begin{equation}
H_{v}=-\widetilde{r}\sum_{(ij)}\sigma _{j}^{x}\sigma
_{i}^{x}-W\sum_{i}\sigma _{i}^{z}  \label{H_V}
\end{equation}
where operators $\mathbf{\sigma} _{i}$ act in the usual way on the states
with/without half-vortices at plaquette $i$ and the first sum runs over
adjacent plaquettes $(ij)$. This Hamiltnian describes an Ising model in a
transverse field. For small $W/\widetilde{r}<\lambda_{c}\sim 1$ its ground
state is "disordered":  $\langle \sigma ^{z}\rangle =0$ but 
$\langle \sigma ^{x}\rangle \neq 0$ while for $W/\widetilde{%
r}>\lambda_{c}$ it is "ordered": $\langle \sigma ^{z}\rangle \neq 0$,
$\langle \sigma ^{x}\rangle =0$. The critical value of transverse field 
is known from extensive numerical simulations \cite{Bishop2000}:
$\lambda_{c}\approx 4.6 \pm 0.3$ for triangular lattice. 
The ''disordered'' state corresponds to the 
liquid of half-vortices, while in the ''ordered'' state the density of
free half-vortices vanishes, i.e. the ground state contains even number of
half-vortices so the total vorticity of the system is zero. To prove this we
start from the state $|\uparrow \rangle $ which is the ground state at 
$\widetilde{r}/W=0$ and consider the effect of $\widetilde{r}%
\sum_{(ij)}\sigma _{j}^{x}\sigma _{i}^{x}$ in perturbation theory. Higher
energy states are separated from the ground state by the gap $W$ so each
order is finite. Further in each order operator $\sigma _{j}^{x}\sigma
_{i}^{x}$ create two more half-vortices proving that the total number of
half-vortices remains even in each order. The
states with odd number of half-vortices have a
gap $\Delta (\widetilde{r}/W)$ which remains non-zero for $W/%
\widetilde{r}>\lambda_{c}$.

In terms of the original discrete variable defined on the rhombi the
Hamiltonian (\ref{H_V}) becomes
\begin{equation}
H_{u}=-\widetilde{r}\sum_{(ij)}\tau _{ij}^{x}-W\sum_{i}\prod_{j(i)}
\tau_{ij}^{z}  \label{H_u}
\end{equation}
where operators act on the state of each rhombus. 
This Hamiltonian commutes with the constraint (\ref{Gauge_inv}) and is in
fact the simplest Hamiltonian of the lattice ${\cal Z}_2$ gauge theory. 
The disordered regime corresponds to a confined phase of this ${\cal Z}_2$
gauge theory, leading to elementary $4e$ charge excitations and the ordered 
regime to the deconfined phase.

Consider now the system with non-trivial topology, e.g. a hole. In this case
the set of variables $v_{i}$ is not sufficient to determine uniquely the
state of the system, one has to supplement it by the variable $%
v_{0}=\sum_{L}u_{ij}$ where sum is taken over a closed contour $L$ that goes
around the hole. Physically, it describes the presence/absence of the
half-vortex in the hole. The effective Hamiltonian of this additional
variable has only kinetic part because presence or absence of half vortex in
a hole which has $l$ weak links in its perimeter gives potential energy $%
W_{0}=c\epsilon _{J}\left( \frac{\epsilon _{J}}{\epsilon _{C}}\right) ^{l}$
which is exponentially small for $l\gg 1$. The kinetic part is similar to
other variables:
$
H_{0}=-\widetilde{r}\sum_{i \in I}\sigma _{i}^{x}\sigma _{0}^{x}, 
$
it describes a process in which half-vortex jumps from the hole into the
inner boundary, $I$, of the system. In the state with 
$\langle \sigma ^{z}\rangle \neq 0$ this process
increases the energy of the system by $\tilde{W} ( \frac{W}{\widetilde{r}} )$ 
($\tilde{W}(0)=W$ and $\tilde{W}(\lambda_c)=0$). 
In the state with $\langle \sigma ^{x}\rangle \neq 0$ it costs nothing. 
Thus, the process in which half-vortex jumps from the hole into the system 
and another half-vortex exits into the outside region appears in the second 
order of the perturbation theory, the amplitude of this process is
$
t_{v}=\widetilde{r}^{2} \sum_{i \in I, j \in O} g_{ij} 
$
where sum is performed over all sites of the inner ($I$) and outer ($O$)
boundaries and $g_{ij}=\left\langle \sigma _{i}^{x}\frac{1}{H-E_{0}}\sigma
_{j}^{x}\right\rangle$ has a physical meaning of the half vortex tunneling
amplitude from inner to outer boundaries. At small $\widetilde{r}/W$, 
we can estimate $g_{ij}$ using the perturbation 
expansion in $\widetilde{r}/W$: the leading contribution appears in $|i-j|$-th 
order of the perturbation theory, thus $g_{ij} \propto 
(\widetilde{r}/W)^{|i-j|}$. Thus for small $\widetilde{r}/W$
the tunneling amplitude of the half vortex is exponentially small in the 
distance, $L$, from the outer to the inner boundary;
we expect that it remains exponentially small for all 
$\widetilde{r}/W < \lambda_c$.  
For $\widetilde{r}/W> \lambda_c $ this amplitude 
is of the order of $\widetilde{r}^{2}/W$ and therefore is significant.

In a different language, in the system with a hole we can construct a
topological invariant $\mathcal{P}=\prod_{\gamma}\tau_{ij}^x$ 
(contour $\gamma$ is shown in Fig.~1)
which can take values $\pm 1$ . The same arguments as used for
the superconducting array show that any dynamics consistent with constraint (%
\ref{Gauge_inv}) preserves $\mathcal{P}$.
Thus, formally, the properties of the topological insulator are very similar
to the properties of the topological superconductor discussed in \cite
{Ioffe2002a}, if one replaces the words Cooper pair by half-vortex and vice
versa. We summarize this duality in the following table

\begin{widetext}
\begin{tabular}{lll}
& \textbf{Topological Superconductor} & \textbf{Topological Insulator} \\ 
Ground state & Condensate of $4e$ charges & Condensate of $2\pi $ phase vortices \\ 
Fluxons & Gapful, charge $2e$ & Gapful, $\pi $ phase vortices \\ 
Pseudocharges & Half fluxes with energy $\epsilon \sim E_J \log (L)$ 
& Charge $2e$ with $\epsilon =2 r$. 
\footnote{
We neglect here the contributions proportional to $\epsilon_C$:
$\delta \epsilon \sim \epsilon_C \min \left[ \log L, \log (c/c_0) \right]$
where $c$ is the capacitance of a weak link and $c_0$ is the
self-capacitance of an island. 
}
\\ 
Ground state degeneracy & Charge on the inner boundary mod $4e$ 
& Number of $\pi $ vortices inside the hole mod $2$ \\ 
Ground state splitting &  
$(\frac{\delta \Phi }{\Phi _{0}}\frac{E_{J}}{r})^L$ 
& $(\widetilde{r}/W)^{L}$
\end{tabular}
\end{widetext}

Note that at small $\widetilde{r}/W\rightarrow 0$ the ground state of the
Hamiltonian (\ref{H_u}) satisfies the condition (\ref{Gauge_inv}) and
minimizes the second term in (\ref{H_u}), i.e. satisfies the condition $%
\prod_{j(i)}\tau_{ij}^{z}|\Psi \rangle =|\Psi \rangle $; it can be
explicitly written as $|0\rangle =$ $\prod_{i}\frac{1}{2}(1+\prod_{j(i)}%
\tau_{ij}^{z})\prod_{kl}|\rightarrow \rangle_{kl}$. This state is a linear
superposition $\frac{1}{\sqrt{2}}(|+\rangle + |-\rangle)$ of the 
degenerates states with $\mathcal{P}=1$ and $\mathcal{P}=-1$; it coincides 
with the ground state of discrete variables in the superconducting array. 
The orthogonal superposition of  $\mathcal{P}= \pm 1$ states,
$\frac{1}{\sqrt{2}}(|+\rangle - |-\rangle)$, corresponds to the half-vortex 
inside the hole and has a much larger energy in the superconductive array. 

The degenerate ground states in the insulating array can be manipulated in
the same way as in the superconductor. Since physically these states
correspond to the absence or presence of the half-vortex inside the hole the
adiabatic change of local magnetic field that drags one half vortex across the
system, flips the state of the system, providing us with the implementation
of the operator $\tau ^{x}$ acting on the state of the qubit. 
Analogously, motion of elementary charge $2e$
around the hole changes the relative phase of the states with and without
half-vortex providing us with the operator $\tau ^{z}$.

The signature of the topological insulator is the persistence of the trapped 
half flux inside the central hole (see Fig.~1)  which can be observed by
cycling magnetic field so as to drive the system back and forth
between insulating and superconducting states. This trapping is
especially striking in the insulator. 
Experimentally, this can be revealed by driving slowly the array into a 
superconducting state and then measuring the phase difference between
opposite points such as $A$ and $B$ in Fig.~1. In the state with a half
vortex the phase difference is $\pi/2 +\pi n$ while it is $\pi n$ in the
other state. The $\pi n$ contribution is due to the usual vortices that
get trapped in a big hole. 
This slow transformation can be achieved by changing the
strength of weak links using the external magnetic field as a control
parameter. The precise nature of the superconductive state
is not essential because phase
difference $\pi$ between points $A$ and $B$ can be interpreted
as due to a full vortex trapped in a hole in a converntional
superconductor or due to a $\pi$ periodicity in a topological one
which makes no essential difference.  
These flux trapping experiments are similar to the ones proposed for
high-$T_c$ cuprates \cite{Senthil2001,Senthil2001b}with a number of 
important differences: the trapped flux is half of $\Phi_0$, the 
cycling does not involve temperature (avoiding problems with 
excitations) and the final state can be either conventional or
topological superconductor. 

We acknowledge the support by NSF grant 4-21262. MF was supported
by SCOPES, RFBR grant 01-02-17759, NWO,
``Quantum Macrophysics'' of RAS/RMS, 
LI is grateful to ENS for the hospitality.


\begin{thebibliography}{99}

\bibitem{Preskill1998}  J. Preskill, Proc. Roy. Soc. A \textbf{454}, 
385 (1998).

\bibitem{Kitaev1997}  A. Kitaev, preprint 
http://xxx.lanl.gov/abs/quant\--ph/9707021 (1997).

\bibitem{Wen1990}  X.-G. Wen and Q. Niu, Phys. Rev. B \textbf{41}, 
9377 (1990).

\bibitem{Wen1991}  X.-G. Wen, Phys. Rev. B \textbf{44}, 2664 (1991).

\bibitem{Ioffe2002a}  L. B. Ioffe and M. V. Feigel'man, cond-mat/0205186.

\bibitem{Ioffe2002b}  L.B. Ioffe \textit{et al}, Nature {\bf 415}, 503 
(2002).

\bibitem{Senthil2000} T. Senthil and M.P.A. Fisher, Phys. Rev. B \textbf{62}, 
7850 (2000).

\bibitem{Senthil2001} T. Senthil and M.P.A. Fisher, Phys. Rev. B \textbf{63},
134521 (2001).

\bibitem{Sachdev1991} N. Read and S. Sachdev, Phys. Rev. Lett. \textbf{66}, 
1773 (1991).

\bibitem{Moessner2002} R. Moessner, S.L. Sondhi and E. Fradkin,
Phys. Rev. B \textbf{65}, 24504 (2002).

\bibitem{Balents2002} L. Balents, M.P.A. Fisher and S.M. Girvin, 
Phys. Rev. B \textbf{65}, 224412 (2002).

\bibitem{Senthil2002} T. Senthil and O. Motrunich, cond-mat/0201320 (2002).

\bibitem{Misguich2002} G. Misguich, D. Serban and V. Pasquier,
cond-mat/0204428 (2002).

\bibitem{Doucot2002}  B. Dou\c{c}ot and J. Vidal, Phys. Rev. Lett. 
\textbf{88}, 227005 (2002).

\bibitem{Bishop2000} R.F. Bishop, D.J.J. Farrel and M.L. Ristig,
Int. J. of Mod. Phys. B, \textbf{14}, 1517 (2000).

\bibitem{Senthil2001b} T. Senthil and M.P.A. Fisher, Phys. Rev. Lett. 
\textbf{86}, 292 (2001). 

\end{thebibliography}
\end{document}